\documentclass[10pt]{article}

\usepackage{colortbl}
\usepackage{amsmath}
\usepackage{amssymb}
\usepackage{graphics}
\usepackage{float}
\usepackage{placeins}
\usepackage{rotating}
\usepackage{cite}
\usepackage{color}
\usepackage{fancybox}
\usepackage{pstricks}
\usepackage{xcolor}
\usepackage{multicol}
\usepackage{framed}
\usepackage{shadowtext}
\newcommand{\cd}{black!50!red!90!}
\newcommand{\cj}{black!60!green!80!}

\newcommand{\ct}{black!30!blue}
\definecolor{bg}{rgb}{0.90, 0.85, 0.40}

\definecolor{bbb}{rgb}{0.91, 1.0, 1.0}
\definecolor{navy}{rgb}{0,0,.6}
\definecolor{jour}{rgb}{0,0.6,.4}
\definecolor{jbul}{rgb}{0.7,0.,.4}

\def\0{\mbox{\tiny $0$}}
\def\1{\mbox{\tiny $1$}}
\def\2{\mbox{\tiny $2$}}
\def\3{\mbox{\tiny $3$}}
\def\4{\mbox{\tiny $4$}}
\def\5{\mbox{\tiny $5$}}
\def\6{\mbox{\tiny $6$}}
\def\7{\mbox{\tiny $7$}}
\def\8{\mbox{\tiny $8$}}
\def\9{\mbox{\tiny $9$}}

\begin{document}

%%%%%%%%%%%%%%%%%%%%%%%%%%%%%%%%%%%%%%%%%%%%%%%%%%%%%%%%%%%%%%%%%%%%%%%%%%%
%%%%%%%%%%%%%%%%%%%%%%%%%%%%%%%%%%%%%%%%%%%%%%%%%%%%%%%%%%%%%%%%%%%%%%%%%%%

\thispagestyle{empty}
\setcounter{page}{0}

{\Large \bf \color{\cd}
\begin{center}
 \fbox{\colorbox{bg}{  \color{\ct}
\begin{tabular}{c}
\shadowrgb{1,1,1}
\shadowoffset{2.5pt}
\color{navy}
\fontsize{14}{14}\selectfont
\bf
\shadowtext{THE OSCILLATORY BEHAVIOR OF LIGHT}\\
\shadowrgb{1,1,1}
\shadowoffset{2.5pt}
\color{navy}
\fontsize{14}{14}\selectfont
\bf
\shadowtext{IN THE COMPOSITE GOOS-H\"ANCHEN SHIFT}

\end{tabular}
 }}
\end{center}
}

\vspace*{1cm}

{\Large \bf
\begin{center}
\color{\cd} $\boldsymbol{\bullet}$
\color{\cj} Physical Review A \color{\cd} 95 \color{\cj}, 053836-9 (2017)
\color{\cd} $\boldsymbol{\bullet}$
\end{center}
}

\vspace*{1cm}

\begin{center}
\begin{tabular}{cc}
\begin{minipage}[t]{0,55\textwidth}
\vspace*{-0.4cm}
\color{black}
{\bf  \color{\cd} Abstract}.
{\color{black} For incidence in the critical region}, the pro\-pagation of gaussian lasers through {\color{black}{triangular}} dielectric blocks
is characterized by {\color{black} the joint action} of angular deviations and lateral displacements. This mixed effect, known as composite Goos-H\"anchen shift, {\color{black} produces a lateral displacement dependent on the axial coordinate, recently confirmed by a weak measurement experiment.}
We discuss under which conditions {\color{black} this axial lateral displacement, which only  exists for the composite Goos-H\"anchen shift, presents an oscillatory behavior}. This oscillation phenomenon shows a peculiar behavior  of light for {\color{black} critical} incidence and, if experimentally tested, could {\color{black}{stimulate further theore\-tical studies and}}
lead to interesting optical applications.
\end{minipage}
&
{\color{\cd} \fbox{\hspace*{-0.12cm} \color{black} {\colorbox{bg}{
\begin{minipage}[t]{0,4\textwidth}
\vspace*{0.15cm}
{\bf \color{\ct} Manoel P. Ara\'ujo}\\
Institute of  Physics ``Gleb Wataghin''  \\
State University of Campinas (Brazil)\\
{\bf \color{\cd} mparaujo@ifi.unicamp.br}
\hrule
\vspace*{0.15cm}
{\bf \color{\ct} Stefano De Leo}\\
Department of Applied Mathematics\\
State University of Campinas (Brazil)\\
{\bf \color{\cd} deleo@ime.unicamp.br}
\hrule
\vspace*{0.15cm}
{\bf \color{\ct}  Gabriel G. Maia}\\
Institute of  Physics ``Gleb Wataghin''\\
State University of Campinas (Brazil)\\
{\bf \color{\cd} ggm11@ifi.unicamp.br}

\end{minipage}
}}}}
\end{tabular}
\end{center}

\vspace*{2cm}

\begin{center}
{\color{\ct}
{\bf
\begin{tabular}{ll}
I. & INTRODUCTION \\
II. & THE COMPOSITE GOOS-H\"ANCHEN SHIFT \\
III. & TRANSMITTED INTENSITY'S ANALYTICAL EXPRESSION\\
IV. & THE AXIAL OSCILLATORY BEHAVIOR\\
V. &CONCLUSIONS\\
& \\
& \,[\, 12 pages, 4 figures\,]
\end{tabular}
}}
\end{center}

\vspace*{4.5cm}

{\Large
\begin{flushright}
\color{\cd} \fbox{\hspace*{-.2cm}
\colorbox{bg}{
\,\color{\cj}$\boldsymbol{\Sigma}$
\color{\cd}$\boldsymbol{\delta}$
\hspace*{-.1cm}\color{\cj}$\boldsymbol{\Lambda}$
}\hspace*{-.2cm}
}
\end{flushright}
}

\newpage

%%%%%%%%%%%%%%%%%%%%%%%%%%%%%%%%%%%%%%%%%%%%%%%%%%%%%%%%%%%%%%%%%%%%%%%%%%
%%%%%%%%%%%%%%%%%%%%%%%%%%%%%%  SECTION I    %%%%%%%%%%%%%%%%%%%%%%%%%%%%%
%%%%%%%%%%%%%%%%%%%%%%%%%%%%%%%%%%%%%%%%%%%%%%%%%%%%%%%%%%%%%%%%%%%%%%%%%%

\section*{\large \color{\ct} I. INTRODUCTION}

The easiest  way of describing laser propagation through dielectric structures is in terms of ray optics.  The geometrical approach is useful to explain most of the practical applications\cite{born,saleh}. In the extraordinary experiment realized by Goos and H\"anchen in 1947\cite{GH47}, the discovery of a lateral displacement of transverse electric (TE) beams totally reflected by a dielectric/air interface suggested deviations from the laws of geometrical optics and stimulated new studies in looking for them. For {\color{black}{$\theta_{\0}>\theta_{_{\rm cri}}+\,\lambda/{\rm w}_{\0}$, where $\theta_{\0}$ is the incidence angle at the left air/dielectric side of the triangular dielectric block depicted in Fig.\,1,
\begin{equation}
\theta_{_{\rm cri}} = \arcsin \left[\left(1-\sqrt{n^{^{2}}-1}\right)/\sqrt{2}\right]
\end{equation}
($\sin\theta=n\,\sin\psi$, $\psi=\varphi+\pi/4$  and $\sin\varphi_{_{\rm cri}}=1/n$),  $\lambda$ is the wavelength of the beam and ${\rm w}_{\0}$ its minimal waist, the gaussian beam is totally reflected by the lower dielectric/air interface. The Fresnel coefficient is complex and acts on the whole gaussian angular distribution.}} The  complex phase is responsible for the transversal shift of the beam. For $\theta_{\0}\gg\theta_{_{\rm cri}}+\lambda/{\rm w}_{\0}$,  the shift is of the order of $\lambda$. Its derivation, done by using the stationary phase method, appeared for the first time in literature one year after the experiment of Goos-H\"anchen and was proposed  by Artmann\cite{GH48}. He also observed that for transverse magnetic (TM) beams a different lateral displacement occurs. In 1949, this prediction was experimentally confirmed by Goos and H\"anchen\cite{GH49}.
Approaching the critical region, $\theta_{\0}\approx\theta_{_{\rm cri}}+\lambda/{\rm w}_{\0}$ the lateral displacement gains an amplification, passing from $\lambda$ to $\sqrt{\lambda\,{\rm w}_{\0}}$\cite{GHamp1,GHamp2,GHamp3}. In the Artmann region, {\color{black}{$\theta_{\0}\gg\theta_{_{\rm cri}}+\lambda/{\rm w}_{\0}$,}} where the shift is proportional to $\lambda$, amplifications can be obtained by a multiple reflections device, as it is the case in the first  Goos-H\"anchen (GH) experiment\cite{GH47}, or by using the weak measurement technique\cite{WM1,WM2}, as recently done by  Jayaswal, Mistura, and Merano\cite{WMGH}.

For  $\theta_{\0}<\theta_{_{\rm cri}}-\lambda/{\rm w}_{\0}$, {\color{black}{the angular gaussian distribution is  modulated}} by real Fresnel coefficients, the complex phase is lost, and, consequently, the lateral GH shift is not present. Nevertheless, in this incidence region, angular deviations from the laws of ray optics occur. This is due to the fact that while the incident gaussian field,
\begin{equation}
I_{_{\rm INC}} \propto \left|\,\int_{_{-\,\pi/2}}^{^{+\,\pi/2}}
\hspace*{-0.7cm}\mbox{d}\theta\,\,\,g(\theta-\theta_{\0})\,\,\exp\left\{i\,k\,\left[\,
(\theta-\theta_{\0})\,y_{_{\rm LAS}}\, - \,(\theta-\theta_{\0})^{^{2}}z_{_{\rm LAS}}/2\,\right)\,
\right\}\,\right|^{^{2}}\,\,,
\end{equation}
has a symmetric angular distribution,
\[
g(\theta-\theta_{\0}) = \exp[\,-\,(\theta-\theta_{\0})^{^{2}} (k\,{\rm w}_{\0})^{^{2}}/4\,\,,
\]
$k=2\,\pi/\lambda$, centered at $\theta_{\0}$, and consequently moves along the $z_{_{\rm LAS}}$ axis,  the transmitted field (see Fig.\,1) is
\begin{equation}
\label{ITRA}
\hspace*{-.5cm}I_{_{\rm TRA}}^{^{[\alpha]}} \propto \left|\,\int_{_{-\,\pi/2}}^{^{+\,\pi/2}}
\hspace*{-0.7cm}\mbox{d}\theta\,\,\,g(\theta-\theta_{\0})\,\,
L_{_{\alpha}}(\theta)\,D_{_{\alpha}}(\theta)\,R_{_{\alpha}}(\theta)\,\exp\left\{i\,k\,\left[\,
(\theta-\theta_{\0})\,(y-y_{\0})\, - \,(\theta-\theta_{\0})^{^{2}}z/2\,\right)\,
\right\}\,\right|^{^{2}}\,\,,
\end{equation}
where $y_{\0}= (\,\sin\theta_{\0} + \cos\theta_{\0}\,)\,a  + \{\,[\,\cos\theta_{\0} -n\,\cos\psi(\theta_{\0})\,]\,\sin\theta_{\0}\,/\,n\cos\psi(\theta_{\0})\,\}\,b$\,\,
is the geometrical shift\cite{AJP}, $\alpha$ the polarization,
\begin{equation*}
\left\{\,L_{_{\rm TE}}(\theta)\,R_{_{\rm TE}}(\theta)   \,,\,L_{_{\rm TM}}(\theta)\,R_{_{\rm TM}}(\theta) \,\right\}  =
\left\{\,
\frac{ 4\,n\,\cos\theta\,\cos\psi}{[\,\cos\theta + n\,\cos\psi]^{^{2}}}
\,,\,
\frac{ 4\,n\,\cos\theta\,\cos\psi}{[\,n\,\cos\theta + \cos\psi]^{^{2}}}
\,\right\}\,\,,
\end{equation*}
$\sin\theta=n\,\sin\psi$, the transmission Fresnel coefficients at the left and right interfaces, and
\begin{equation}
\label{FreC}
\left\{\,D_{_{\rm TE}}(\theta)\,,\,\, D_{_{\rm TM}}(\theta)\,\right\} = \left\{\,\frac{n\,\cos\varphi - \sqrt{1-n^{^{2}}\sin^{\2}\varphi}}{n\,\cos\varphi + \sqrt{1-n^{^{2}}\sin^{\2}\varphi}}\,\,,\,\, \frac{\cos\varphi -n \sqrt{1-n^{^{2}}\sin^{\2}\varphi}}{\cos\varphi+ n\,\sqrt{1-n^{^{2}}\sin^{\2}\varphi}}\,\right\}\,\,,
\end{equation}
$\varphi=\pi/4 +\psi$, the reflection coefficient at the down interface,  is characterized by an angular distribution whose initial gaussian shape is now {\em distorted} by the Fresnel coefficients. The  symmetry breaking  caused by the Fresnel coefficient\cite{Break1,Break2}, creates an axial dependence of the transversal component  of the transmitted field  and, as a consequence, angular deviations from the optical path predicted  by the laws of geometrical optics. These deviations are of the order of $(\lambda/{\rm w}_{\0})^{^{2}}$. For transverse magnetic waves and incidence in the vicinity  of the Brewster angle,
the angular deviations gain an amplification of ${\rm w}_{\0}/\lambda$ leading to the giant GH angular shift\cite{ANG1}. This  amplification has recently been  detected  in direct\cite{Exp1} and weak measurement technique-based\cite{Exp2,Exp3} experiments.

{\color{black}{The GH lateral displacement has also recently been investigated in the reflection of a light beam
by a graphene layer and controlled by a voltage modulation\cite{AR1} and in the reflection of terahertz radiation from an uniaxial antiferromagnetic crystal, where the action of an external magnetic field induces a nonreciprocity in the shift for positive and negative incidence\cite{AR2}.
}}

We did not aim to give a complete review of theoretical analyses or experimental facts. We confined ourselves to outline the general field  in which their investigation is seated. For the reader  who wishes to deepen  any question of lateral displacements, angular deviations and/or breaking of symmetry in its entirety, we suggest to read the excellent works of Bliokh and Aiello\cite{Rev1} and  G\"otte, Shinohara, and Hentschel\cite{Rev2},  where clear presentations, detailed discussions, and  relevant aspects of deviations from the laws of ray optics are reported.

{\color{black}{
In the next section, we give a brief discussion of the motivations which stimulated our investigation in the critical region of incidence,
\begin{equation}
\label{criReg}
\theta_{_{\rm cri}}-\,\lambda/{\rm w}_{\0}\, <\, \theta_{\0} \,< \, \theta_{_{\rm cri}}+\,\lambda/{\rm w}_{\0}\,\,,
 \end{equation}
{\color{black} where angular deviations and lateral displacements act together generating the composite GH shift}.  This discussion will then be followed (section III)}}  by an analytical description of the beam transmitted through a triangular dielectric block and by the  calculation of the transversal displacement of the transmitted beam as a function of its axial coordinate $z$. The analytical approximation is then tested and confirmed,{\color{black}{ in section IV}}, by the numerical calculation done by using directly Eq.(\ref{ITRA}). The surprising  result of oscillations in the lateral displacement is, probably, the most important evidence of  the strange behavior of light near the critical incidence. {\color{black}{Conclusion and outlooks are drawn in the final section.}}

\section*{\large \color{\ct} II. THE COMPOSITE GOOS-H\"ANCHEN SHIFT}

The critical region, {\color{black}{see Eq.\,(\ref{criReg}),}}  surely represents the most interesting incidence region to study deviations from the laws of ray optics. This is due to the fact that, in this region, angular deviations and lateral displacements  are both present. Furthermore, at critical incidence,
 the breaking of symmetry is maximized\cite{Break1,Break2} and the Goos-H\"anchen shift  amplified by a factor $\sqrt{k\,{\rm w}_{\0}}$\cite{GHamp1,GHamp2,GHamp3}. In such a region, we thus expect amplified angular deviations. Numerical calculations, done in this region,  show  a clear amplified axial dependence of the GH shift\cite{Axial}. These (numerical theoretical) predictions have recently been confirmed by a weak measurement experimental analysis\cite{CGH}.  Once confirmed the amplified axial dependence, it would be important  to understand if, and if so, under which conditions {\color{black}{negative}} lateral displacements occur. The idea of light oscillations  {\color{black}{around the path predicted by the geometrical optics, for incidence}} in the critical region,  was stimulated by the fact that in this region  the gaussian angular distribution is modulated by a reflection Fresnel coefficient which is in part real ($\theta<\theta_{_{\rm cri}}$) and in part complex ($\theta>\theta_{_{\rm cri}}$). The modulated angular distribution, which is responsible for the spatial shape of the transmitted field (no longer a gaussian field), should  present an interference between the real and complex part generating oscillation phenomena.

The possibility to find an analytical expression for the transmitted field allows to calculate the maximal transmitted intensity and to determine the  beam parameters for which oscillations can be experimentally detected. Clearly, once obtained the analytical approximation, the analytical expression  has to be tested by numerical calculations done by using directly the transmitted intensity given in Eq.\,(\ref{ITRA}).  Numerical calculations are often hard,
require time and leave obscure how to optimize the choice of the beam parameters or the experimental device. So, the analytical expression proposed in {\color{black}{the next section}} could be very useful for experimentalists interested to study, investigate and, possibly, detect, in the critical region, angular deviations from the law of geometrical optics and oscillation phenomena.

{\color{black}{
Analytical expressions for the GH shift, extensively studied in literature, always represented an intriguing  challenge. The first attempt was carried out by Artmann in 1948\cite{GH48}. Its derivation explained the lateral displacement of TE waves found in the experiment realized by Goos and H\"anchen  one year before\cite{GH47} and predicted a different behavior of TM waves, later confirmed by Goos and H\"anchen in the experiment of 1949\cite{GH49}. Nevertheless, the Artmann formula contains a divergence at the critical angle and this is due to the fact that the Taylor expansion used for the complex phase breaks down when the incidence angle approaches the critical one. In 1971\cite{GHamp1}, Horowitz and Tamir (HT) proposed an analytical expression for the lateral displacement of  a gaussian light beam incident from a denser to a rarer medium. They obtained an
approximation for the Fresnel coefficient which allowed to analytically solve the integral determining
the propagation of the reflected beam and found for the TE and TM lateral displacements, a closed expression in terms of parabolic-cylinder (Weber) functions. They also found normalized curves valid for a wide range of parameters and suggested that the general functional behavior of the lateral shift should be similar for other symmetric angular distributions. In 1986\cite{AR3}, Lai, Cheng, and Tang (LCT) overcame the cusp-like structure in the HT formula obtaining a theoretical result for the lateral shift of a gaussian light beam which varies continuously and smoothly around the critical angle. Recently\cite{GHamp3}, a closed form expression for the GH lateral displacement was proposed by Ara\'ujo, De Leo, and Maia (ADM). The ADM formula, differently from the HT and LCT ones, is not based on the reflection coefficient expansion but on the integral analysis of the complex phase. In ref.\,\cite{GHamp3}, the analytical expression obtained for the lateral displacement of gaussian light beam neglecting the axial dependence is given in terms of modified Bessel functions of the first kind. The analysis done in\cite{GHamp3} is also extended to different angular distribution shapes  and also distinguishes between the lateral displacement of the optical beam maximum and the mean valued calculation of its shift which, due to the angular breaking of symmetry in the critical region, are different. {\color{black}{ The HT\cite{GHamp1}, LCT\cite{AR3}, and ADM\cite{GHamp3} formulas reproduce for $\theta_{\0}\gtrsim \theta_{_{\rm cri}}+\lambda/{\rm w}_{\0}$ the Artmann prediction and overcome for incidence at critical angle the
infinity problem. Such formulas are obtained for $z\ll z_{_{\rm R}} (=\pi {\rm w}_{\0}^{^{2}}/\lambda)$
and  do \underline{not} contain any axial dependence.} This means, for example, that to experimentally reproduce the theoretical results given in ref. \cite{GHamp1,GHamp3,AR3}  the camera  has to be moved very close to the (right) interface of the  triangular dielectric block}. Axial dependence requires a more complicated study  and often numerical calculations\cite{Axial}. Its effect, which also appears before the critical region  leading to angular deviations, produces, in the critical region, an {\color{black}{axial}} amplification of the lateral displacement {\color{black}{with respect to the amplification proportional to $\sqrt{k\,{\rm w}_{\0}}$}}  predicted by the HT, LCT, and ADM formulas. This amplification  has recently been  seen in the weak measurement experiment cited in ref.\cite{CGH}. In the next section, we will find an analytical expression of the transmitted intensity without any axial simplification. So, our final formula will explicitly contain the $z$-dependence of the camera position {\color{black} coming from  the $z$-depending term of the spatial phase appearing in Eq.\,(\ref{ITRA}), i.e.}
\[  {\color{black}{k\,(\,\theta-\theta_{\0}\,)^{^{2}}\,z\,/2\,=\,}}[\,k\,{\rm w}_{\0}\,(\,\theta-\theta_{\0}\,)\,]^{^{2}}\,z\,/z_{_{\rm R}}\,\,.\]
{\color{black} If a gaussian beam, incident upon a dielectric/air  interface (like the lower interface of the triangular prism), has its incidence angle in the critical region and if its angular distribution is broad enough
(${\rm w}_{\0}\ll 1\,{\rm mm}$), plane waves in the angular spectrum with $\theta>\theta_{_{\rm cri}}$
will be totally internally reflected and  plane waves with $\theta<\theta_{_{\rm cri}}$
partially reflected. The angular breaking of symmetry and the real ($\theta<\theta_{_{\rm cri}}$) and complex ($\theta>\theta_{_{\rm cri}}$) nature of the reflection coefficient  for incidence in the critical region play a fundamental role  in the the oscillatory behavior of the lateral displacement seen in  the composite GH shift.    It is the interplay between the Goos-H\"anchen shift (total internal reflection) and the angular deviations (partial reflection) which generates  the composite Goos-H\"anchen effect.  The $z$-dependence of the lateral displacement in the critical region \cite{Axial} has recently been experimentally confirmed by using the weak measurement technique in ref.\,\cite{CGH}. In this paper, we analyze under which conditions an oscillatory behavior occurs and when the pattern of oscillation can be reproduced by wider beams.}

The analysis presented in this paper applies to coherent light fields and leads to an analytical formula in terms of confluent hypergeometric functions of first kind. Partially coherent light fields have to be treated by using the Mercer expansion as done in ref.\,\cite{AR4}.}}

\section*{\large \color{\ct} III. TRANSMITTED INTENSITY'S  ANALYTICAL EXPRESSION}

To obtain an analytical formula for the transmitted beam, some approximations have to be done. The first approximation is to factorize the left/right transmission coefficients. Such  coefficients are very smooth varying functions in the critical region and can thus be calculated in $\theta_{\0}$. The second approximation is to change  the limits of integration from $\pm \,\pi/2$ to $\pm\, \infty$. This is possible as our incident gaussian is strongly centered in $\theta_{\0}$ which varies between $\theta_{_{\rm cri}}-\lambda/{\rm w}_{\0}$ and $\theta_{_{\rm cri}}+\lambda/{\rm w}_{\0}$ (for BK7 prism $\theta_{_{\rm cri}}=-\,5.603^{^{o}}$). Without loss of generality, we can thus rewrite the transmitted intensity as follows
\begin{equation}
\label{TRAnum}
I_{_{\rm TRA}}^{^{[\alpha]}} \propto \left|\,\int_{_{-\,\infty}}^{^{+\,\infty}}
\hspace*{-0.5cm}\mbox{d}\theta\,\,\,
D_{_{\alpha}}(\theta)\,\exp\left\{\,-\,\left[\,\frac{k\,{\rm w}_{\0}}{2}\,(\theta-\theta_{\0})\,\zeta(z)\,\right]^{^{2}}+\,i\,k\,{\rm w}_{\0}\,(\theta-\theta_{\0})\,\delta_{_{\rm GH}}\,\right\}
\,\right|^{^{2}}\,\,,
\end{equation}
where $\delta_{_{\rm GH}}=(y-y_{\0})/{\rm w}_{\0}$ and $\zeta(z)=\sqrt{1+2\,i\,z/k\,{\rm w}_{\0}^{^{2}}}$.
The crucial point in the analytical approximation is to develop  the reflection coefficient $D_{_{\alpha}}(\theta)$  in square-root powers around the critical angle. {\color{black} To do it, we rewrite the incidence angle $\theta$ in terms of the critical one, $\theta_{_{\rm cri}}$,
\[\theta = \theta-\theta_{_{\rm cri}} + \theta_{_{\rm cri}}=\delta\theta + \theta_{_{\rm cri}}\,\,.\]
By observing that
\[\sin\theta=n\,\sin\psi\,\,\,\,\,\Rightarrow\,\,\,\,\,\delta\theta\,\cos\theta_{_{\rm cri}} =n\,\delta \psi \cos \psi_{_{\rm cri}}\,\,\,\,\,\,\,{\rm and}\,\,\,\,\,\,\,
 \psi=\varphi+\pi/4\,\,\,\,\,\Rightarrow\,\,\,\,\,\delta\psi=\delta\varphi\,\,,\]
 we obtain
 \[\delta\varphi =\frac{\cos\theta_{_{\rm cri}}}{\cos\psi_{_{\rm cri}}}\,\frac{\delta \theta}{n}\approx \frac{\delta \theta}{n}\,\,,\]
 for  BK7 $\varphi_{_{\rm cri}}=\arcsin(1/n)=41.305^{^{o}}$,  $\psi_{_{\rm cri}}=-\,3.695^{^{o}}$ and
 $\theta_{_{\rm cri}}=-\,5.603^{^{o}}$.  By expanding  around the critical angle
 \begin{eqnarray*}
 n\,\cos \varphi &\approx &  n\,\cos \varphi_{_{\rm cri}} - n\, \sin \varphi_{_{\rm cri}}\,\delta\varphi \,\,=\,\,  n\,\cos \varphi_{_{\rm cri}} - \delta\varphi\,\,,\\
 \sqrt{1-n^{^2}\sin^{^{2}}\varphi} & \approx & \sqrt{1-n^{^2}\sin^{^{2}}\varphi_{_{\rm cri}}-\,2\,n^{^{2}}\sin\varphi_{_{\rm cri}}\,\cos\varphi_{_{\rm cri}}\,\delta\varphi } \,\,=\,\, \sqrt{-\,2\,n\,\cos\varphi_{_{\rm cri}}\,\delta\varphi }\,\,,
 \end{eqnarray*}
 and introducing the quantity $\delta\phi = - \delta\varphi/n\,\cos\psi_{_{\rm cri}}$, we can approximate   the TE and TM  reflection coefficients, given in (\ref{FreC}), as follows
 \begin{eqnarray*}
 D_{_{\rm TE}} & = & \frac{1+\delta\phi - \sqrt{2\,\delta\phi}}{ 1+\delta\phi + \sqrt{2\,\delta\phi}} \,\,\approx\,\, 1 -\,2 \,\sqrt{2\,\delta\phi}  -\,4\,\delta \phi\,\,,  \\
 D_{_{\rm TM}} & = & \frac{1+\delta\phi - n^{^{2}}\,\sqrt{2\,\delta\phi}}{ 1+\delta\phi + n^{^{2}}\,\sqrt{2\,\delta\phi}} \,\,\approx\,\, 1 -\,2 \,n^{^{2}}\,\sqrt{2\,\delta\phi}  -\,4\,n^{^{2}}\,\delta \phi\,\,.
 \end{eqnarray*}
 Finally,
}
\begin{equation}
D_{_{\alpha}}(\theta) \,\approx\,
1
- \sqrt{\,2\,\gamma_{_{\alpha}}\,(\theta_{_{\rm cri}} -\theta)\,}
+ \gamma_{_{\alpha}}\,(\theta_{_{\rm cri}}-\theta)
\,\,,
\end{equation}
where $\left\{\,\gamma_{_{\rm TE}}\,,\,\gamma_{_{\rm TM}}\,\right\} =
4\,\{\,1\,,\,n^{^{4}}\,\}\,/\,n\,\sqrt{n^{^{2}}-1}$. By using this expansion and introducing the new integration variable
$\tau=k\,{\rm w}_{\0}\,\zeta(z)\,(\theta_{_{\rm cri}}-\theta)/2$, we obtain
\begin{eqnarray}
\label{ITRAs}
I_{_{\rm TRA}}^{^{[\alpha]}} & \propto &  \left|\,
\int_{_{-\,\infty}}^{^{+\,\infty}}
\hspace*{-0.5cm}\mbox{d}\tau\,\,\,\left[\,1 - 2\, \sqrt{\frac{\gamma_{_{\alpha}}\,\tau}{k\,{\rm w}_{\0}\,\zeta(z)}} + \frac{2\,\gamma_{_{\alpha}}\,\tau}{k\,{\rm w}_{\0}\,\zeta(z)} \,\,\right]\,\,\, \exp\{\,-\,[\,
\tau + \,d(\delta_{_{\rm GH}},z)\,]^{^{2}}\,\}\,\right|^{^{2}}\mathcal{G}(\delta_{_{\rm GH}},z)\,\,,
\end{eqnarray}
where
\[d(\delta_{_{\rm GH}},z;\theta_{\0})= k\,{\rm w}_{\0}\,\zeta(z)\,(\theta_{\0}-\theta_{_{\rm cri}})/2 + i \,\delta_{_{\rm GH}}/\zeta(z)\]
contains, besides the transversal and axial variables,  the incidence angle dependence and
  $\mathcal{G}(\delta_{_{\rm GH}},z)$ represents  the gaussian function $\exp[-\,2\,\delta_{_{\rm GH}}^{^{2}}/|\zeta(z)|^{^{4}}]$. The square-root and linear terms in the brackets of Eq.\,(\ref{ITRAs}) act as modulators of  the gaussian function   $\mathcal{G}(\delta_{_{\rm GH}},z)$. The effect of their modulation can be evaluated once the integral in  Eq.\,(\ref{ITRAs}) is analytically solved. To do this, we observe that the integral with a square-root integrand can be converted into a series
 \begin{eqnarray}
 \int_{_{-\,\infty}}^{^{+\,\infty}}
\hspace*{-0.5cm}\mbox{d}\tau\,\,\,\sqrt{\tau\,}\,\,\exp[\,-\,(\,\tau+d\,)^{^2}\,] & = & \exp[\,-\,d^{^{2}}\,]\,\sum_{_{m=0}}^{^{\infty}}\,\frac{(-\,2\,d)^{^{m}}}{m!}\, \int_{_{-\,\infty}}^{^{+\,\infty}}
\hspace*{-0.5cm}\mbox{d}\tau\,\,\,\tau^{^{m+\frac{1}{2}}}\,\,\exp[\,-\,\tau^{^2}\,] \nonumber\\
 & = & \exp[\,-\,d^{^{2}}\,]\,\sum_{_{m=0}}^{^{\infty}}\,\frac{(2\,d)^{^{m}}}{m!}\, \frac{(-1)^{^{m}}+1}{2}\,\,\Gamma \left(\,\frac{2\,m+3}{4}\,\right)\,\,.
 \end{eqnarray}
The series found is a well-known  Gamma functions series leading to a combination of confluent hypergeometric functions of
 first kind. Thus, the integral can be analytically solved in terms of these functions
  \begin{equation}
 \int_{_{-\,\infty}}^{^{+\,\infty}}
\hspace*{-0.5cm}\mbox{d}\tau\,\,\,\sqrt{\tau\,}\,\,\exp[\,-\,(\,\tau+d\,)^{^2}\,]  =   \frac{e^{i\,\frac{\pi}{4}}}{2\,\sqrt{2}}\,\,\mathcal{F}(\delta_{_{\rm GH}},z;\theta_{\0})
 \end{equation}
 where
 \[ \mathcal{F}(\delta_{_{\rm GH}},z;\theta_{\0}) =  \left[\,2 \,\Gamma \left(\mbox{$\frac{3}{4}$}\right) \, _1F_1\left(\mbox{$\frac{3}{4},\,\frac{1}{2},\,d^{^2}$}\right)+\,i\, \Gamma \left(\mbox{$\frac{1}{4}$}\right) \, _1F_1\left(\mbox{$\frac{5}{4},\,\frac{3}{2},\,d^{^2}$}\right)\,d\right]\,\exp[\,-\,d^{^2}\,]\,\,.
 \]
 Finally, the analytical approximation for the transmitted intensity is given by
 \begin{eqnarray}
\label{ITRAs2}
I_{_{\rm TRA}}^{^{[\alpha]}} & \propto &  \left|\,
1 -  \sqrt{\frac{\gamma_{_{\alpha}}}{2\,\pi\,k\,{\rm w}_{\0}\,\zeta(z)}}\,e^{i\,\frac{\pi}{4}}\,\mathcal{F}(\delta_{_{\rm GH}},z;\theta_{\0})    - \frac{2\,\gamma_{_{\alpha}}\,d(\delta_{_{\rm GH}},z;\theta_{\0})}{k\,{\rm w}_{\0}\,\zeta(z)} \,\,\right|^{^{2}} \mathcal{G}(\delta_{_{\rm GH}},z)\,\,.
\end{eqnarray}
{\color{black} The study of the maximum of this function will be the topic of the next section.}

\section*{\large \color{\ct} IV. THE AXIAL OSCILLATORY BEHAVIOR}
{\color{black}{
In order to check the validity of our approximation and compare our results with the previous ones appeared in literature, we first analyze the case in which the axial dependence is removed from Eq.\,(\ref{ITRAs2}). In the experimental setup  this means the case in which the camera is positioned very close to the right side of the dielectric block, i.e. $z\ll z_{_{\rm R}}$. This is for example the situation  of the experiment done by Cowan and Anicin in ref.\cite{Conw}. In such an experiment, the collected data were compared with the theoretical formulas of Artmann\cite{GH48} and  Tamir and Horowitz \cite{GHamp1}. In Fig.\,2a, we plot, for TE waves, the lateral displacement for a laser   gaussian beam with wavelength $\lambda=0.633$
(the wavelength of choice for most HeNe lasers) and  beam waists of $150,\,300,\,{\rm and}\,600\,\,\mu{\rm m}$.
transmitted through a triangular BK7 ($n=1.515$) block,  and detected by a camera very close to the right side of the dielectric block. As observed, this means to remove the axial dependence in Eq.\,(\ref{ITRAs2}). In this case, we obtain, in accordance with the previous theoretical calculations appeared in literature \cite{GHamp1,GHamp3,AR3},  an amplification of the lateral shift proportional to $\sqrt{k\,{\rm w_{\0}}}$ for critical incidence, see Fig.\,2a. In this approximation, the amplification  prefers wider spatial distributions.
The numerical calculations, obtained by using directly  Eq.\,(\ref{TRAnum}) with $\zeta(z)\approx 1$ are in excellent agreement with the results predicted by the analytical formula  Eq.\,(\ref{ITRAs2}) for $z\approx 0$.
Phenomena as angular deviations and/or oscillations cannot be seen in this case.
 By moving the camera away from the right side of the dielectric block, along the axial propagation direction of the transmitted beam predicted by geometric optics, an additional $z$-dependent lateral displacement  appears. This is for example the case of the experimental setup in the ref.\,\cite{Muller}. This $z$-dependent lateral displacement  is  called in literature angular shift. This angular shift   is clearly visible in the left incidence region of Fig.\,2b-c. The axial dependence for critical region mixes two effects: the angular deviations (caused by the symmetry breaking of the angular distribution) and the GH shift (caused by the additional complex phase in the Fresnel coefficient). This mixed effect, known as composite GH shift, was recently proven by a weak measurement experiment \cite{CGH}.  The axial effects depend on the ratio $z/z_{_{\rm R}}$ and consequently, for a fixed axial position of the camera, narrower spatial beams experience larger amplifications. This can be understood by observing that narrower spatial beams have  wider angular distributions and consequently they are more sensible to the breaking of symmetry caused by the Fresnel reflection coefficient. This axial amplification, different from the standard amplification obtained for $z\approx 0$,  is clearly seen in the plots of Figs.\,2b-c. In such plots, we also see {\em new} oscillation phenomena for ${\rm w}_{\0}=150\,\mu{\rm m}$. The numerical data show  an excellent agreement with the analytical calculation.
The axial amplification was recently confirmed by the experiment done in ref.\,\cite{CGH}.
Nevertheless, the oscillatory behavior was not detected in such an experiment because, as seen in Fig.\,2b-c,
 the oscillatory behavior starts,  for a beam waist of $150\,\mu{\rm m}$,  from an axial position of the camera of $25$ cm and it is seen for incidence greater than the critical ones.  In the referred experimental article, the beam waist was of $170\,\mu{\rm m}$, the camera positioned at $z=20$ and $25$ cm and the incidence angle not great enough to reach the right zone of the critical region. The mathematical explanation  of the oscillation
phenomenon comes from the presence of confluent hypergeometric functions in the transmitted intensity. For $\theta_{\0}\leq\theta_{_{\rm cri}}$ (the dominant part of $d$ is the real one)  and $\theta_{\0}\geq\theta_{_{\rm cri}}+\lambda/{\rm w}_{\0}$ (the dominant part of $d$ is the imaginary one), the argument of
the  confluent hypergeometric functions is real and no oscillation can be seen. In the right zone of the critical region, $\theta_{_{\rm cri}}\leq \theta_{\0}\geq\theta_{_{\rm cri}}+\lambda/{\rm w}_{\0}$, depending on the value of $z/z_{_{\rm R}}$ the real and complex parts of $d$ become comparable and  the confluent hypergeometric functions will have a complex argument opening the door to oscillation phenomena. For increasing values of the incidence angle the beam reaches the Artmann zone and the angular distribution recovers its original  gaussian symmetry leading to the Artmann results. In this incidence region,  the composite GH shift tends to the standard GH shift which only depends on the geometrical dielectric structure and on the beam wavelength and, consequently, the shift for different beam waists is the same for different beam waists, see the right zone of the critical incidence in Figs.\,2.
}}

The amplification  $\sqrt{\gamma_{_{\rm TM}}/\gamma_{_{\rm TE}}}=n^{^{2}}$  between TE and TM waves is of a factor 2.3  (BK7 prism) and it is shown in Figs.\,3a-b. The amplification between the different beam waist $150\,\mu{\rm m}$ (Fig.\,3c) and $600\,\mu{\rm m}$  (Fig.\,3d) is of a factor $\sqrt{{\rm w}_{_{600}}/{\rm w}_{_{150}}}=2$.  The fact that the axial coordinate $z$ always appears in the analytical formula with the denominator $k\,{\rm w}_{{\0}}^{^{2}}$ also allows to predict when, given two different  gaussian beam waists, it is possible to visualize the same pattern of oscillation. For the cases examined in our study,   the  axial coordinate multiplication factor is given by  $z_{_{600}}/ z_{_{150}} =  ({\rm w}_{_{600}}/{\rm w}_{_{156}})^{^{2}}=16$. According to the results shown in Fig.\,3, it seems that by increasing the beam waist we improve our experimental measurement. In reality, what we improve  is the lateral displacement of a factor   $\sqrt{{\rm w}_{_{600}}/{\rm w}_{_{150}}}=2$. Nevertheless, what we measure is the lateral displacement of a beam with waist ${\rm w}(z)$.
Thus, it would be  more appropriate to introduce an adimensional quantity given by the ratio between the lateral displacement and the beam waist at the axial point where the camera is located. This ratio assesses the experimental performance needed  to measure the lateral displacement. For the cases examined in Fig.\,3c and 3d, we have $\sqrt{\lambda/{\rm w}_{_{150}}}$  and $\sqrt{\lambda/{\rm w}_{_{600}}}$ respectively, clearly showing a better efficiency for a measurement done with a beam waist of $150\,\mu{\rm m}$.

The analysis presented in Fig.\,4 has been carried out  by calculating the lateral displacement as a function of the incidence angle $\theta_{\0}$ for different axial position of the camera. It is also interesting to calculate  such a displacement as a function of the axial position $z$ for fixed incidence angles. Approaching the critical angle from the left (see Fig.\,4a), we have the phenomenon of amplification of the angular deviations (caused by the angular symmetry breaking of  the transmitted
beam), i.e. $0.0011^{^{o}}$ ($\theta_{\0}=-5.8^{^{o}}$), $0.0017^{^{o}}$ ($-5.7^{^{o}}$), and $0.0055^{^{o}}$ ($-5.6^{^{o}}$). For incidence greater that the critical angle, the oscillating GH shift appears, see Fig.\,4(c-f).   Fixing the  incidence  angle at $\theta_{_{\0}}=-5.54^{^{o}}$, and varying the axial position of the camera at $z=20,\,40,\,70,\,100$\,cm, we would find a pattern of oscillating GH displacement, $7.6,\,0.4\,-\,16.3,\,30.0\,\mu{\rm m}$.  The amplitude of oscillation decreases  when the incidence angle approaches the right border of the critical region.

\section*{\large \color{\ct} V. CONCLUSIONS}

Finally, we can conclude that, in the critical region, for $\theta_{\0} <\theta_{_{\rm cri}}$ the real part of the angular distribution wins over the complex one and angular deviations are the main evidence of the angular breaking of symmetry. For  $\theta_{\0} > \theta_{_{\rm cri}}$, the situation is reversed and the main contribution comes from the complex part generating oscillation phenomena. By increasing the incidence angle, and approaching the right border of the critical region, the real part of the angular distribution vanishes and we recover the gaussian symmetry in the transmitted beam. In this case, we do not find any notable angular deviations. The beam practically moves along the $z$-axis but it is transversally displaced by the GH shift. For incidence at $\theta_{\0}=-\,5.4^{^{o}}$,  and different axial positions of the camera, say $20,\,40,\,70,\,100$\,cm,  we, for example, find the following lateral displacements  $2.6,\,2.7,\,2.8,\,2.9\,\mu{\rm m}$.

To conclude this work let us emphasize once more that the challenge of detecting deviations from the laws of geometrical optics is still a current issue of optics containing a number of unsolved and, at the same time, interesting questions of very general significance. We have not given a rigorous mathematical elaboration of the theory but only a simplified analytical formula to calculate angular deviations and oscillation phenomena in the critical region.   It is the authors hope that this study will find many readers among theoretical and experimental physicists and specialists in related branches of optics, by helping them in future theoretical studies as well as stimulating experiments that could confirm the oscillatory Goos-H\"anchen shift.

\section*{\normalsize \color{\ct} ACKNOWLEDGEMENTS.}
{\color{black}{ In deep gratefulness an appreciation, the authors would like to thank the referees and editor for their attentive reading, their willingness to discuss, their suggestions, and their challenging questions. They belong to what is still a small but fortunately growing minority of revisors: people who are able to delve deeply into the article and with their stimulating questions allow to improve its scientific content. Some additional questions were asked by the editors in order to provide further clarification of certain critical points. The paper in its present form would not have come into existence without their support.}}

\newpage

\newpage

\begin{figure}[ht]
	\centering
	\vspace{-2.2cm}
		\hspace*{-1.3cm}
		\includegraphics[scale=1]{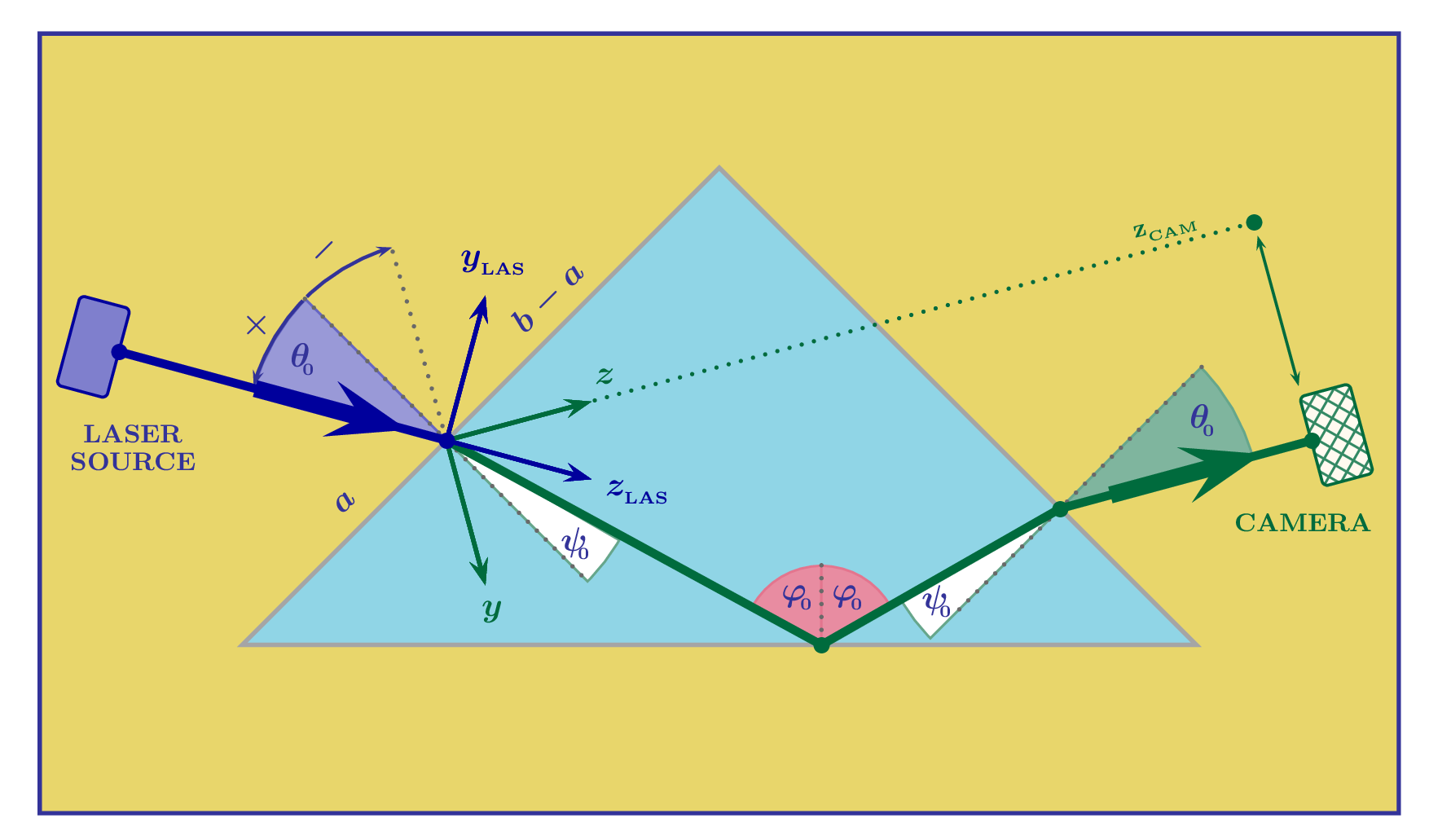}
		\vspace{-0.5cm}
		\caption{\color{black} \rm  The geometrical set-up of the experiment on detecting oscillations. The incoming Gaussian beam propagates along the $z_{_{\rm INC}}$ axis forming an angle $\theta_{\0}$ (the incidence angle) with the normal to the left (air/dielectric) prism interface. Its minimal beam waist is found at the point in which the beam is refracted by the left interface. After the first refraction ($\psi_{\0}$), the beam is reflected ($\varphi_{\0}$) by the down (dielectric/air) interface and, finally, refracted ($\theta_{\0}$) by the right one, reaching the camera positioned at an axial distance $z_{_{\rm CAM}}$ from the point of minimal beam waist.}
\end{figure}

\newpage

\begin{figure}[ht]
% 	\centering
	\vspace{-2.2cm}
		\hspace{-1.3cm}
		\includegraphics[height=22.5cm, width=18cm]{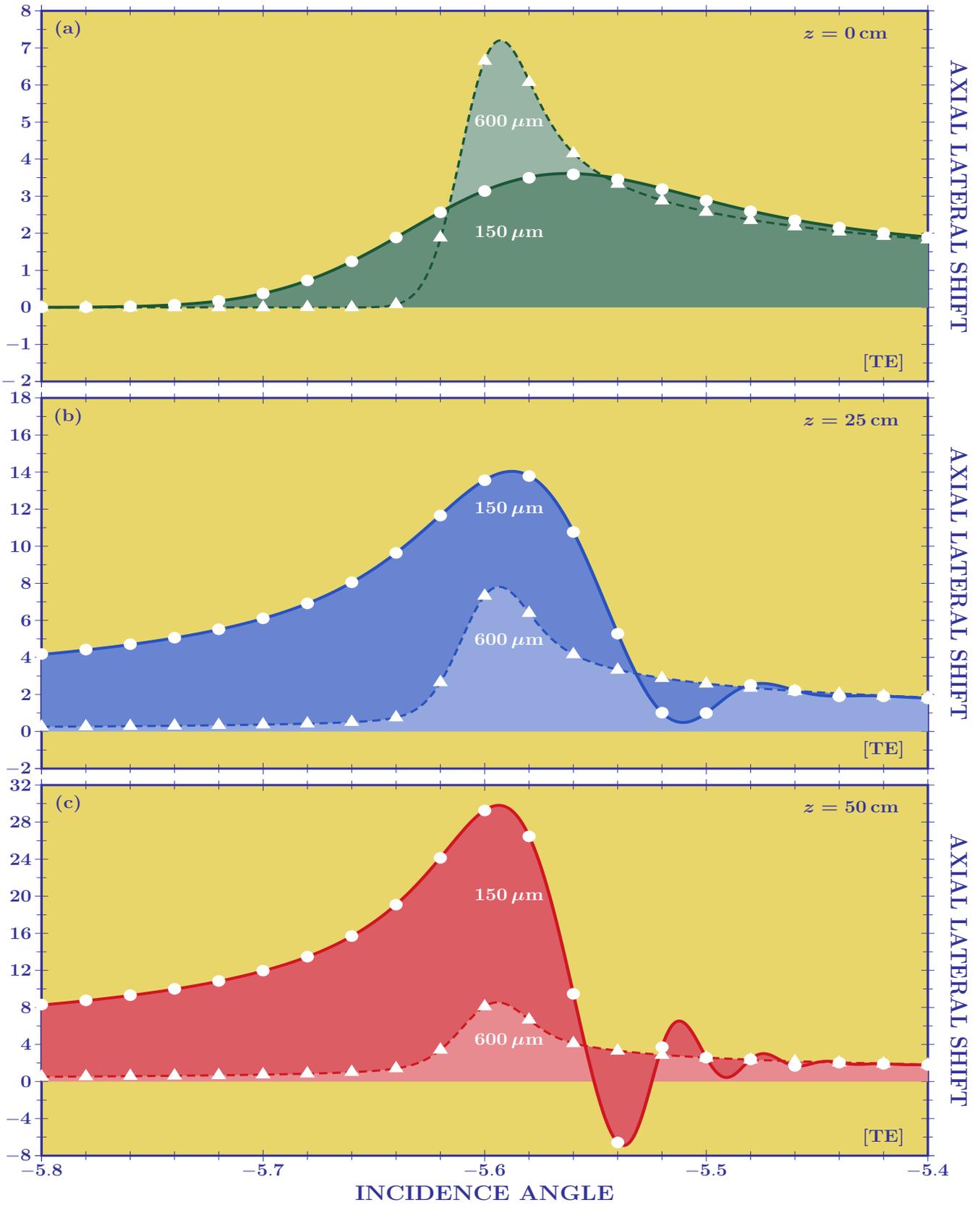}
	\vspace{-0.5cm}
		\caption{\color{black} \rm  Lateral displacement, as a function of the incidence angle, for TE waves.
Dashed lines represent the analytical results for beams with a ${\rm w}_{\0}=600\,\mu m$ while solid lines are used for beams with ${\rm w}_{\0}=150\,\mu m$. The white circles ($150\,\mu m$) and white triangles  ($600\,\mu m$) indicate the numerical data showing  an excellent agreement with the  analytical results. In (a) the camera is positioned very close to the right face of the prism, in (b) at $z=25\,{\rm}cm$, and in (c) at $z=50\,{\rm cm}$.  In (a), the results of literature are recovered. Nevertheless,
as the camera moves away from the prism, an axial dependence appears, favoring beams with wider angular distributions. As $z$ increases, we can also see new oscillation phenomena.}
\end{figure}

\newpage

\begin{figure}[ht]
% 	\centering
	\vspace{-2.2cm}
		\hspace{-1.3cm}
		\includegraphics[height=22.5cm, width=18cm]{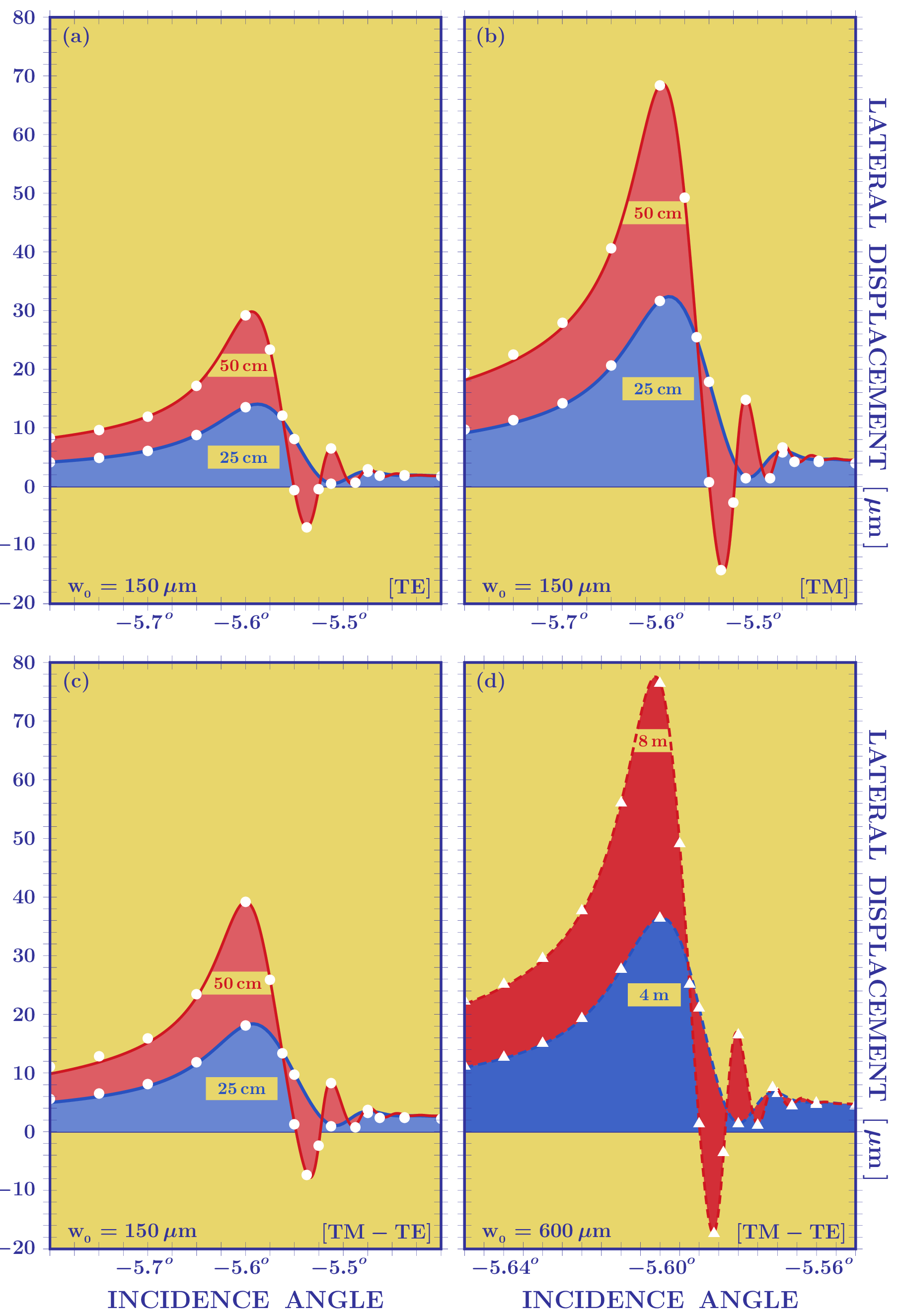}
		\vspace{-0.5cm}
		\caption{\color{black} \rm  In (a) and (b), the lateral displacement for an optical beam with  ${\rm w}_{\0}=150\,\mu{\rm m}$ is shown for TE and TM waves respectively. For a BK7 prism the amplification factor is $2.3$. In (c) and (d), the lateral displacement is plotted for two different beam waists, $150$ and $600\,\mu{\rm m}$. For an axial distance amplified by $({\rm w}_{_{600}}/{\rm w}_{_{150}})^{^{2}}$ and an incidence region reduced by ${\rm w}_{_{150}}/{\rm w}_{_{600}}$, we recover in (d) the same oscillation pattern of (c). The numerical calculations (circles and triangles) show an excellent agreement with the analytical results (continuous and dashed lines). {\color{black} As expected the axial dependence breaks down when the incidence angle approaches the Artmann zone where the stationary method works fine.}}
\end{figure}

\newpage

\begin{figure}[ht]
% 	\centering
	\vspace{-2.2cm}
		\hspace{-1.3cm}
		\includegraphics[height=22.5cm, width=18cm]{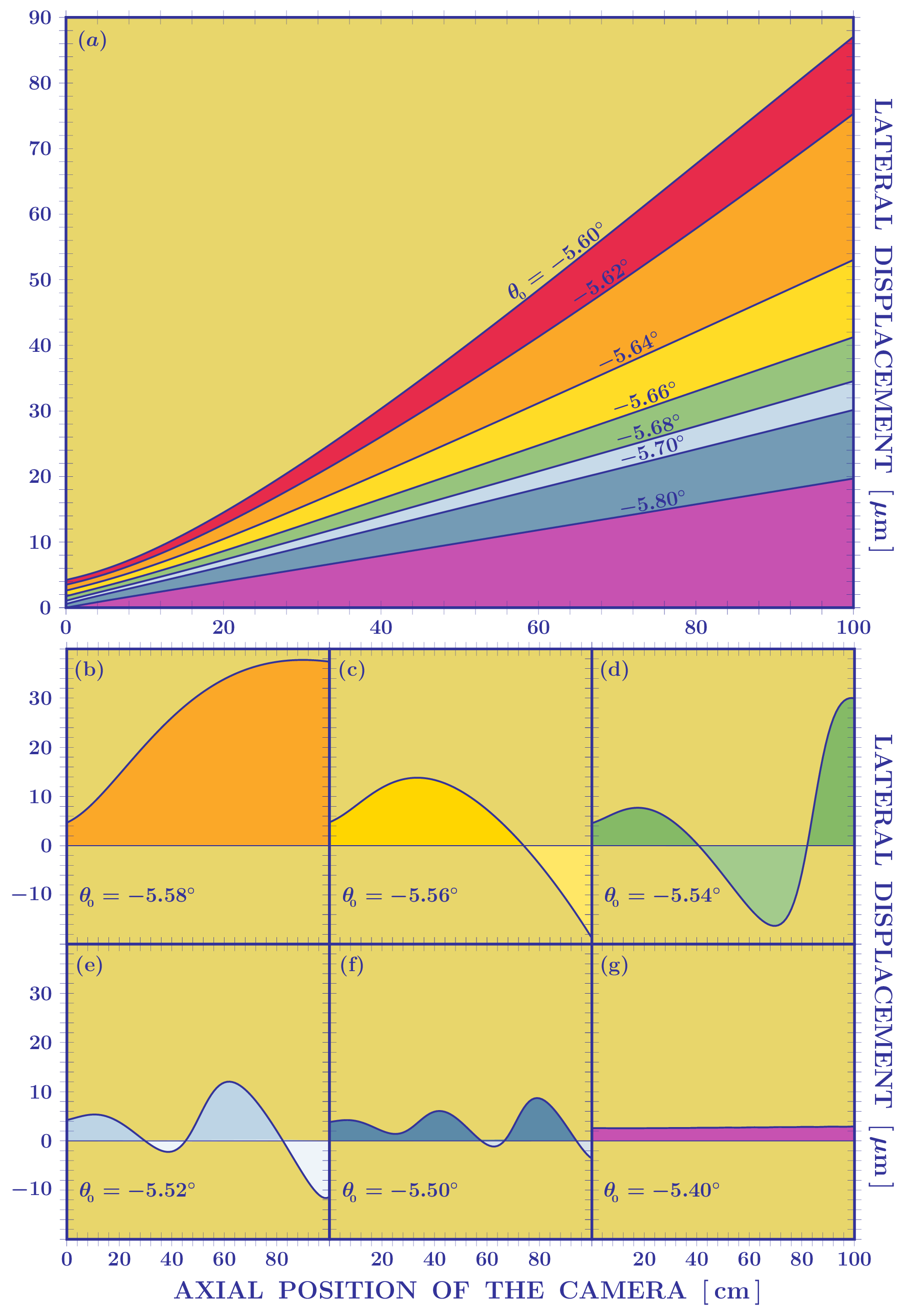}
		\vspace{-0.5cm}
		\caption{\color{black} \rm    Lateral displacement as a function of the axial position of the camera. In (a), the phenomenon
of angular deviations as well as its amplification for incidence angles approaching the critical one is evident. In (b-f), oscillations are visible. Their amplitude is reduced when the incidence angle approaches the left border of the critical region. In (g), the symmetry of the angular distribution is completely recovered and angular deviations as well as oscillations are lost.}
\label{fig3}
\end{figure}

\end{document}